# Practical tests of neutron transmission imaging with a superconducting kinetic-inductance sensor

The Dang Vu[a,e], Hiroaki Shishido[b,c], Kazuya Aizawa[a], Kenji M. Kojima[d,e],
Tomio Koyama[e], Kenichi Oikawa[a], Masahide Harada[a], Takayuki Oku[a], Kazuhiko Soyama[a],
Shigeyuki Miyajima[f], Mutsuo Hidaka[g], Soh Y. Suzuki[h], Manobu M. Tanaka[i], Alex Malins[j],
Masahiko Machida[j], Shuichi Kawamata[c,e] and Takekazu Ishida[c,e,*]

[a]*Materials and Life Science Division, J-PARC Center, Japan Atomic Energy Agency (JAEA), Tokai, Ibaraki 319-1195, Japan*

[b]*Department of Physics and Electronics, Graduate School of Engineering, Osaka Prefecture University, Sakai, Osaka 599-8531, Japan*

[c]*NanoSquare Research Institute, Osaka Prefecture University, Sakai, Osaka 599-8570, Japan*

[d]*Centre for Molecular and Materials Science, TRIUMF, 4004 Wesbrook Mall, Vancouver, BC V6T 2A3, Canada*

[e]*Division of Quantum and Radiation Engineering, Osaka Prefecture University, Sakai, Osaka 599-8570, Japan*

[f]*Advanced ICT Research Institute, National Institute of Information and Communications Technology (NICT), Kobe, Hyogo 651-2492, Japan*

[g]*Advanced Industrial Science and Technology (AIST), Tsukuba, Ibaraki 305-8568, Japan*

[h]*Computing Research Center, Applied Research Laboratory, High Energy Accelerator Research Organization (KEK), Tsukuba, Ibaraki 305-0801, Japan*

[i]*Institute of Particle and Nuclear Studies, High Energy Accelerator Research Organization (KEK), Tsukuba, Ibaraki 305-0801, Japan*

[j]*Center for Computational Science & e-Systems, Japan Atomic Energy Agency (JAEA), 178-4-4 Wakashiba, Kashiwa, Chiba 277-0871, Japan*

* E-mail: ishida@center.osakafu-u.ac.jp

**ABSTRACT**

Samples were examined using a superconducting (Nb) neutron imaging system employing a delay-line technique which in previous studies was shown to have high spatial resolution. We found excellent correspondence between neutron transmission and scanning electron microscope (SEM) images of Gd islands with sizes between 15 and 130 μm which were thermally-sprayed onto a Si substrate. Neutron transmission images could be used to identify tiny voids in a thermally-sprayed continuous $Gd_2O_3$ film on a Si substrate which could not be seen in SEM images. We also found that neutron transmission images revealed pattern formations, mosaic features and co-existing dendritic phases in Wood's metal samples with constituent elements Bi, Pb, Sn and Cd. These results demonstrate the merits of the current-biased kinetic inductance detector (CB-KID) system for practical studies in materials science. Moreover, we found that operating the detector at a more optimal temperature (7.9 K) appreciably improved the effective detection efficiency when compared to previous studies conducted at 4 K. This is because the effective size of hot-spots in the superconducting meanderline planes increases with temperature, which makes particle detections more likely.

## 1. Introduction

Neutron beams are sensitive not only to certain light elements, such as hydrogen, lithium, boron, carbon and oxygen, but also to exceptionally heavy elements with high neutron absorption cross-sections, such as gadolinium, samarium, europium and cadmium. The characteristics of experiments conducted using neutron beams are thus remarkably different from those using X-ray, electron or proton beams. Neutron imaging has been used for taking transmission spectra [1,2], for neutron tomography [3], and as a non-destructive technique for investigating pore structures in materials, for example, a pore structure was observed in the attenuation and dark-field images of an electron beam-melted Ti-6Al-4V cube. [4].

A spatial resolution for neutron imaging of 20 μm was reported using a scintillator camera detector, which decreased to 2 μm when using a CMOS sensor for center-of-gravity corrections [5]. A spatial resolution of 5.4 μm was reported for color-center formation in LiF crystals [6]. A $^{10}$B-doped multichannel plate [7] was used for imaging with pulsed neutron sources with a resolution of 55 μm (or 10 μm with center-of-gravity corrections). Our group developed the current-biased kinetic inductance detector (CB-KID) and demonstrated resolutions of 22 μm [8] and 16 μm [9]. Thin-film-coated thermal neutron detectors with high efficiency have also been created using microstructured semiconductor neutron detector (MSND) technology [10].

In most cases these systems were characterized using a test sample with a fabricated pattern such as a Gd based Siemens star [11] or an array of $^{10}$B dots [8,9]. In reality, however, test samples of interest in material sciences tend to have wide size, thickness, shape and composition distributions. The purpose of this study was to test the practicality of the CB-KID system for imaging samples with more realistic features. The samples studied include Gd and $Gd_2O_3$ films deposited on Si substrates, and Wood's metal samples containing irregular eutectic phases. Wood's metals are of interest to material scientists studying the mechanism of the formation of patterns during irregular eutectic solidification [12,13].

## 2. Methods: Using CB-KID for neutron transmission imaging

The CB-KID system was proposed as a superconducting neutron detector. Its operating principle is somewhat similar to a superconducting single photon detector (SSPD) [14], but in contrast to SSPD, CB-KID works even under a small bias current. A transient change in the density of electrons in the superconducting wire $n_s$ occurs at a hot spot created by a passing charged particle. Although CB-KID was proposed for neutron imaging, it may be used for detecting hot spots created by other stimuli. London-Maxwell theory [15] predicts that a negative pulse propagates in the downstream direction from the hot spot, while a positive pulse propagates upstream. A superconductor-insulator-superconductor planar structure in CB-KID provides an efficient waveguide to transmit electromagnetic pulse signals [15]. When used for neutron imaging, a $^{10}$B conversion layer is needed to convert neutrons to charged particles, which in turn create hot spots in the meanderlines.

In **Fig. 1**, we show a schematic diagram of the CB-KID measurement system. The CB-KID used in this study was fabricated on Si as (1) a 300 nm thick Nb plane, (2) a 350 nm SiO$_2$ layer, (3) a 50 nm thick $Y$ meanderline, (4) a 150 nm SiO$_2$ layer, (5) a 50 nm thick $X$ meanderline, and (6) a 150 nm SiO$_2$ layer. A $^{10}$B neutron conversion layer of 70 nm thickness was deposited on top of the CB-KID by electron-beam evaporation in an ultra-high vacuum chamber. A thin $^{10}$B layer was used for this study as we are still optimizing the process of depositing $^{10}$B to obtain a uniform layer. In past experiments we have encountered issues with $^{10}$B peeling off during deposition, however we expect process optimizations in future will enable us to deposit thicker $^{10}$B conversion layers while ensuring that the detector does not heat up too much causing damage. The 0.9 μm wide $X$ and $Y$ meanderlines had 10,000 repetitions of segments ($h$ = 15.1 mm) with a pitch $p$ = 1.5 μm, giving a sensitive area of 15×15 mm. DC bias currents (50 μA) were fed into the $X$ and $Y$ meanderlines. A 32 Ch time-to-digital convertor with 1 ns sampling (Kalliope-DC readout circuit) received four positive signals [8,16]. The CB-KID temperature was controlled at 7.9 K. The coordinates ($X,Y$) of hot spots were estimated as $X = (t_{Ch4} - t_{Ch3})v_x p/2h$ and $Y = (t_{Ch2} - t_{Ch1})v_y p/2h$, where $t_{Ch1}$, $t_{Ch2}$, $t_{Ch3}$, and $t_{Ch4}$ are the signal arrival times. We measured the signal propagation velocities along the meanderlines as $v_x = 6.052 \times 10^7$ m/s and $v_y = 4.581 \times 10^7$ m/s at 7.9 K by feeding a pulse signal from one end of each meanderline to the other. Neutron images were rendered from the hot spot distribution.

We prepared test samples of Gd islands and Gd$_2$O$_3$ films by thermal spray coating onto a 0.75 mm thick Si substrate (4×4 mm). The Gd islands have a wide distribution of thicknesses. Most islands were less than 2 μm thick, while a few islands had thicknesses as high as ~4 μm. The thickness of the continuous Gd$_2$O$_3$ film was approximately 19 μm.

During etching two 50 μm stainless steel (type 304) masks with 100 μm stripe patterns (pitch 250 μm) were superimposed onto the Si substrate with a small overlapping angle (~7 degrees). The two overlapping masks created lamellar moiré patterns with a repetition pitch of ~2 mm at the open parts.

Wood's metal samples were prepared as buttons from liquid Bi (50%), Pb (26.7%), Sn (13.3%) and Cd (10%). Wood's metal is known to form various different eutectic microstructures during the solidification process [17]. Six samples of Wood's metal were sliced from the buttons using a diamond saw. The sample thickness varied between 0.2 and 0.8 mm in this work.

Each of the test samples was fixed on an Al plate using epoxy and placed at 0.8 mm distance from the CB-KID meanderlines. This was to minimize smearing of the resulting images arising from the angular beam divergence of the pulsed neutron beam.

## 3. Results and discussion
### 3.1. Observation of Gd islands

**Fig. 2(a)** shows a neutron transmission image of the Gd sample acquired over 138 hours with 0.1 to 1.13 nm wavelength pulsed neutrons at 520 kW and a collimation ratio of *L/D*=140 (14 m collimator length and 0.1×0.1 m moderator) at BL10, J-PARC [18]. Note that the beam collimator was fully open so that the whole moderator could be seen from the detector position. Symbolic for many samples of practical interest, there are many islands with varying sizes visible in the neutron transmission image. A scanning electron microscope (SEM) image of the same sample (**Fig. 2(b)**) also shows many islands (white regions) with various sizes. We confirmed that these white islands contained Gd using energy dispersive X-ray spectroscopy. We attribute the stripe pattern visible in **Fig. 2(b)** to re-deposition of stainless steel (type 304) from the masks which occurred in the milling chamber during preparation of the test sample. This was confirmed by energy dispersive X-ray spectroscopy (EDX) analyses, which showed appreciable amounts of Fe and Cr in the stripe locations. These stripes are not visible in the neutron transmission image (**Fig. 2(a)**) due to the low neutron absorption cross sections of Fe and Cr.

The dotted areas highlighted in **Fig. 2(a)** and **Fig. 2(b)** are enlarged in **Fig. 3(a)** and **Fig. 3(b)**, respectively, to show the correspondence between the neutron transmission and SEM images. **Fig. 3(c)** and **Fig. 3(e)** show line profiles in the neutron transmission image along the *x* and *y* directions of the marked Gd island, respectively. The line profiles were fitted with

$$I(x) = I_0 + A \left[ \tanh\left(\frac{x-x_1}{x_w}\right) - \tanh\left(\frac{x-x_1-x_s}{x_w}\right) \right] \qquad (1)$$

where $I_0$ is the floor intensity, $A$ is an amplitude of the peak (or trough), $x_1$ is the position of the peak (or trough), $x_w$ is the width of the island edge and $x_s$ is a measure of the size of the island. This function is convenient for modelling the various shapes of the line profiles across islands, as it can fit both narrow and wide peaks (troughs), and peaks (troughs) with sharp or flat summits (minima). In the analyses below, the full size of a Gd island was calculated as including the edge widths, i.e. $D_x = x_s + 2x_w$. In **Fig. 3(d)** and **Fig. 3(f)**, we show the line profiles in the SEM image of the Gd island along the *x* direction and the *y* direction, respectively. There was good agreement between the sizes from neutron and SEM images: $D_x^{\text{neutron}} = 78.7 \pm 1.5$ μm, $D_x^{\text{SEM}} = 74.2 \pm 1.3$ μm, $D_y^{\text{neutron}} = 78.3 \pm 1.1$ μm and $D_y^{\text{SEM}} = 76.7 \pm 1.7$ μm. We noticed a visible tail at the bottom of the marked island both in the transmission image of **Fig. 3(a)** as well as in the SEM image of **Fig. 3(b)**. We evaluated the width of the tail as $D_x^{\text{neutron}} = 22.3 \pm 8.0$ μm and $D_x^{\text{SEM}} = 21.9 \pm 0.7$ μm. Dotted ovals are shown around two Gd islands in the transmission image of **Fig. 3(a)**. These islands cannot be discerned in the SEM image in **Fig. 3(b)** as they lie within one of the gray stripes.

**Fig. 3(g)** shows the correlation between $D^{\text{neutron}}$ and $D^{\text{SEM}}$ calculated along the *x* and *y* directions for multiple islands on the sample. The data points are fitted linearly by $D^{\text{neutron}} = a\, D^{\text{SEM}}$. We found $a_x = 1.11 \pm 0.02$ for the *x* direction and $a_y = 1.13 \pm 0.03$ for the *y* direction, for islands ranging in size from 15 to 130 μm. The fact that $a_x$ and $a_y$ are different from unity is not because of aberrations in the neutron transmission images, but due to the thicknesses of the Gd islands tapering to zero towards the island edges. SEM is more sensitive to the thin Gd at the edges of the islands than neutron transmission imaging which requires a certain thickness of Gd to give good contrast. If aberrations were the cause, we would expect that the sensitive area of the CB-KID, i.e. the area covered by the $^{10}$B conversion layer (15×15 mm), would be inaccurate when calculated from the neutron transmission image. In fact measuring from the neutron transmission image gives 15.002×15.054 mm, which is close to the true value.

### 3.2. Voids in Gd₂O₃ film

Based on the above positive results, we took a neutron transmission image of the Gd₂O₃ film prepared by thermal-spray coating. The detection of pores using neutron imaging has been studied in the past for electron beam-melted Ti-6Al-4V [3]. **Fig. 4(a)** shows a neutron transmission image of the Gd$_2$O$_3$ sample taken with neutron wavelengths (0.1 to 1.13 nm). Although the image is rather continuous, tiny orange colored bright spots are visible. **Fig. 4(b)** shows a SEM image of the Gd$_2$O$_3$ sample, and there is no indication of these bright spots. We therefore infer that the Gd$_2$O$_3$ film is continuous rather than having isolated island-like structures like the Gd sample, and the orange spots in **Fig. 4(a)** represent voids in the film.

To investigate the void sizes, the dotted area highlighted in **Fig. 4(a)** containing one bright spot is enlarged in **Fig. 4(c)**. There is clearly a void at the center identified by the green region in **Fig. 4(c)**. Fitting the line profiles along the *x* and *y* directions in the transmission image of the void with **Eq. (1)** gives $D_x = 54.8 \pm 9.3$ µm and $D_y = 31.0 \pm 6.7$ µm (**Fig. 4(d)** and **(e)**). The fine morphology of the Gd$_2$O$_3$ film on a 30 µm scale is apparent over the $600 \times 600$ µm enlarged area in **Fig. 4(c)**. This fine morphology is not as clear in **Fig. 4(a)** because the scale for the intensity covers a wider range. The observed morphology indicates the existence of subtle heterogeneities in the thermally-sprayed Gd$_2$O$_3$ film. Neutron transmission imaging is thus a practical technique for checking the internal structures of films for defects that cannot be seen with optical microscopes or SEM imaging.

*3.3. Pattern formation in Wood's metal*

There are four different phases in Wood's metal, one of which is a Cd-rich phase that tends to have a needle-like structure [19,20]. This phase is accessible with neutron transmission imaging due to the large neutron absorption cross section of Cd. The other three constituent elements of Wood's metal (Bi, Pb, Sn) are almost transparent to neutron beams due to their small absorption cross sections. An earlier work used neutron imaging to reveal needle-like precipitates of cadmium in the microstructure of a Wood's metal sample [21].

**Figs. 5(a)-(f)** show the CB-KID neutron transmission images (0.1 to 1.13 nm) of Wood's metal samples. The images reveal various interesting structures including dendritic structures, seen at $(x, y) \simeq (0 \text{ µm}, 1.3 \times 10^3 \text{ µm})$ in **Fig. 5(b)** and at $(x, y) \simeq (4 \times 10^3 \text{ µm}, 0.8 \times 10^3 \text{ µm})$ in **Fig. 5(d)**. Based on differences in the transmission

intensities for different neutron wavelengths, we identified that the bright dendritic lines are from Cd-rich phases, while the surrounding dark lines visible in the images are from Cd-deficient phases [19,20]. Microstructures can also be seen in the images, for example, the lamella pattern in **Fig. 5(a)** near $(x, y) \simeq (-1.7 \times 10^3 \mu m, 4.3 \times 10^3 \mu m)$. This pattern has dark stripes of $\sim 30$ μm width, and a repetition pitch of $\sim 50$ μm.

The microstructures in the samples were not visible using a conventional optical microscope. SEM can probe the surface of samples, but these samples were too thick (0.2 to 0.8 mm) for SEM to image these microstructures which are inside the samples. The thick horizontal and vertical white lines in **Fig. 5(a)**, **(b)** and **(c)** are not genuine but artifacts from defects in the CB-KID. The sensitivity of the CB-KID deteriorates in the segments near the defect of the meanderline, yielding the types of white lines that are visible. One way to remove the effect of the defects is to normalize against an image that is obtained without setting a sample. We demonstrated this in a preceding publication [9]. In the present study, we did not apply open beam normalization because our samples and the detector were installed at a cryogenic temperature in a high vacuum (see **Fig. 1**). Resetting the components to take an image without a sample for normalization would have reduced the beam time available for other measurements for this study.

The thin periodic vertical stripes in **Fig. 5** are also artifacts. The data processing technique yields discrete rather than continues coordinates (*X,Y*) for hot spots due to the repetitive structure of the meanderline, which has a period of $p = 1.5$ μm. This means some pixels have a different effective size to others, giving rise to the periodic anomalies visible in **Fig. 5**. Nonetheless, the present results demonstrate that neutron transmission imaging can be used to study the pattern formations from irregular eutectic solidification of Wood's metal samples.

*3.4. Detection efficiency of CB-KID*

We previously reported the detection efficiency of the *X* meanderline in the CB-KID at 4 K [22]. The delay-line method requires the detection of four signals arriving over a short timescale from the *X* and *Y* meanderlines to identify the coordinate of each hot-spot. For 0.025 eV thermal neutrons, the ratio of efficiencies for simultaneous *X* and *Y* detections compared to detections on the *Y* meanderline was just 12% when the detector temperature was 4 K [23]. In **Fig. 6**, experimental detection efficiencies of single *X* or *Y* meanderline detections and simultaneous *X* and *Y* meanderline detections are shown as a function of neutron wavelength (or time of flight for travelling 14 m) when the detector temperature was controlled at 7.9 K. We found that the ratio of efficiencies for 0.025 eV thermal neutrons for simultaneous *X* and *Y* detections to *Y* detections increased to 82% for the detector at 7.9 K. Although the absolute values of the efficiencies in **Fig. 6** are low, we note that the thickness of the $^{10}$B neutron conversion layer can be increased from 70 nm to 1000 nm to improve the efficiency of CB-KID.

The PHITS Monte Carlo particle transport code was previously applied for estimating the detection efficiency of CB-KID with a 10 μm thick $^{10}$B conversion layer [24]. For comparison with the experimental data in **Fig. 6**, new PHITS simulations were carried out with a 70 nm thick $^{10}$B conversion layer for the present work. The agreement between the experiments and the simulations is fairly good. Some of the discrepancy between the simulation and experimental efficiencies can be explained by the experimental readout circuit (Kalliope-DC) setting a finite threshold for discriminating signals from noise.

We used PHITS simulations to consider the effect of different $^{10}$B conversion layer thicknesses on the detection efficiency. The simulated detection efficiency for 0.025 eV thermal neutrons was 1.4% for a 1 μm thick $^{10}$B conversion layer, increasing to 1.9% for a 2 μm thick $^{10}$B conversion layer. Note that the efficiency is not expected to increase proportionally with the film thickness, as the short ranges of $^{7}$Li particles makes it increasingly unlikely for them to reach the superconducting meanderlines.

We note that the ratio of efficiencies for simultaneous *X* and *Y* detections to *Y* detections was 64% in the simulations for 0.025 eV thermal neutrons. The efficiency of *X* and *Y* simultaneous detections in the PHITS simulations is mainly determined by the spatial coverage of superconducting meanderlines within the detector. The large increase in the ratio of efficiencies between operating the detector at 4 and 7.9 K, and the higher fraction of *X* and *Y* simultaneous detections in the experiments (82%) than the PHITS simulations (64%), may be explained by the hot-spot sizes being larger than the repetition period of the meanderlines when the critical temperature is approached.

Not only increasing the thickness of the conversion layer but also optimizing the operating temperature may be ways to increase the effective detection efficiency of CB-KID. Since the operating temperature of the detector has not yet been optimized to maximize the detection efficiency, it may be possible to achieve higher efficiencies by studying the temperature-dependency of the efficiency systematically.

## 4. Conclusions

We carried out a systematic investigation of the distribution of Gd islands on a thermally-sprayed sample of Gd on a Si substrate by means of neutron transmission imaging and SEM observations. The sizes of the Gd islands determined from the transmission image correlated strongly with those determined from the SEM image. We demonstrated that the CB-KID could be used to identify (1) tiny voids in a thermally-sprayed continuous $Gd_2O_3$ film and (2) various patterns, mosaic morphologies and different eutectic microstructures consisting of Cd-rich phases in Wood's metal samples. The fact that the CB-KID system could be used to probe samples with a wide distribution of sizes and thicknesses is promising for applying the device for practical transmission imaging for samples of interest to material scientists. Operating the CB-KID at higher temperatures appreciably improved the efficiency for simultaneously detecting the *X* and *Y* coordinates of hotspots. Finally we compared the detection efficiency with the PHITS simulations. We now plan to improve the detection efficiency by increasing the thickness of the $^{10}B$ neutron conversion layer.

**Acknowledgments**

This work was partially supported by a Grant-in-Aid for Scientific Research (Grant Nos. JP16H02450, JP19K03751, JP21K14566, JP21H04666) from JSPS. The neutron-irradiation experiments conducted at the Materials and Life Science Experimental Facility (MLF) of the J-PARC were supported by MLF programs (Proposals Nos. 2019A0004, 2019P0201, 2020P0201).

## References

[1]  N. Kardjilov, I. Manke, A. Hilger, M. Strobl, J. Banhart, Advances in neutron imaging,

Mater. Today **14** (6) (2011) 248–256, https://doi.org/10.1016/S1369-7021(11)70139-0.

[2] A. Steuwer, P. J. Withers, J. R. Santisteban, L. Edwards, Using pulsed neutron transmission for crystalline phase imaging and analysis, J. Appl. Phys. **97** (2005) 074903, https://doi.org/10.1063/1.1861144.

[3] R. Woracek, D. Penumadu, N. Kardjilov, A. Hilger, M. Boin, J. Banhart, I. Manke, 3D mapping of crystallographic phase distribution using energy-selective neutron tomography, Adv. Mater. **26** (2014) 4069-4073, https://doi.org/10.1002/adma.201400192

[4] A. J. Brooks, J. Ge, M. M. Kirka, R. R. Dehoff, H. Z. Bilheux, N. Kardjilov, I. Manke, L. G. Butler, Porosity detection in electron beam-melted Ti-6Al-4V using high resolution neutron imaging and grating-based interferometry, Prog. Addit. Manuf. **2** (2017) 125-132, https://doi.org/10.1007/s40964-017-0025-z.

[5] D. S. Hussey, J. M. LaManna, E. Baltic, D. L. Jacobson, Neutron imaging detector with 2μm spatial resolution based on event reconstruction of neutron capture in gadolinium oxysulfide scintillators, Nucl. Instrum. Meth. Phys. Res. A **866** (2017) 9–12, https://doi.org/10.1016/j.nima.2017.05.035.

[6] M. Matsubayashi, A. Faenov, T. Pikuz, Y. Fukuda, Y. Kato, Neutron imaging of micron-size structures by color center formation in LiF crystals, Nucl. Instrum. Phys. Res. A **62**2 (3) (2010) 637-641, https://doi.org/10.1016/j.nima.2010.07.058.

[7] A. S. Tremsin, J. V. Vallerga, J. B. McPhate, O. H. W. Siegmund, Optimization of high count rate event counting detector with Microchannel Plates and quad Timepix readout, Nucl. Instrum. Meth. Phys. Res. A **787** (2015) 20-25, https://doi.org/10.1016/j.nima.2014.10.047.

[8] H. Shishido, Y. Miki, H. Yamaguchi, Y. Iizawa, V. The Dang, K.M. Kojima, T. Koyama, K. Oikawa, M. Harada, S. Miyajima, M. Hidaka, T. Oku, K. Soyama, S.Y. Suzuki, T. Ishida, High-speed neutron imaging using a current-biased delay-line detector of kinetic inductance, Phys. Rev. Appl. **10** (2018) 044044, https://doi.org/10.1103/PhysRevApplied.10.044044.

[9] Y. Iizawa, H. Shishido, K. Nishimura, T. D. Vu, K. M. Kojima, T. Koyama, K. Oikawa, M. Harada, S. Miyajima, M. Hidaka, T. Oku, K. Soyama, K. Aizawa, S. Y. Suzuki, T. Ishida, Supercond. Sci. Technol. **32** (2019) 125009,

https://doi.org/10.1088/1361-6668/ab4e5c.

[10] R. G. Fronk, S. L. Bellinger, L. C. Henson, T. R. Ochs, C. T. Smith, J. K. Shultis, D. S. McGregor, Dual-sided microstructured semiconductor neutron detectors (DSMSNDs), Nucl. Instrum. Meth. Phys. Res. A **804** (2015) 201–206, https://doi.org/10.1016/j.nima.2015.09.040.

[11] P. Trtik, M. Meyer, T. Wehmann, A. Tengattini, D. Atkins, E. H. Lehmann, M. Strobl, PSI 'neutron microscope' at ILL-D50 beamline - first results, Materials Research Proc. **15** (2020) 23-28, https://doi.org/10.21741/9781644900574-4.

[12] A. J. Shahani, X. Xiao, P. W. Voorhees, The mechanism of eutectic growth in highly anisotropic materials, Nature Communications, **7** (2016) 12953, https://doi.org/10.1038/ncomms12953.

[13] S. Akamatsu, M. Plapp, Eutectic and peritectic solidification patterns, Current Opinion in Solid State and Materials Science **20** (2016) 46-54, https://doi.org/10.1016/j.cossms.2015.10.002.

[14] G. N. Gol'tsman, O. Okunev, G. Chulkova, A. Lipatov, A. Semenov, K. Smirnov, B. Voronov, A. Dzardanov, Picosecond superconducting single-photon optical detector, Appl. Phys. Lett. **79** (2001) 705, https://doi.org/10.1063/1.1388868.

[15] T. Koyama, T. Ishida, Electrodynamic theory for the operation principle of a superconducting kinetic inductance stripline detector J. Phys. Conf. Ser. **1054** (2018) 012055, https://doi.org/10.1088/1742-6596/1054/1/012055.

[16] K. M. Kojima, T. Murakami, Y. Takahashi, H. Lee, S. Y. Suzuki, A. Koda, I. Yamauchi, M. Miyazaki, M. Hiraishi, H. Okabe, S. Takeshita, R. Kadono, T. Ito, W. Higemoto, S. Kanda, Y. Fukao, N. Saito, M. Saito, M. Ikeno, T. Uchida, M. M. Tanaka, New µSR spectrometer at J-PARC MUSE based on Kalliope detectors, J. Phys. Conf. Ser. **551** (2014) 012063, https://doi.org/10.1088/1742-6596/551/1/012063.

[17] Y. Saiki, T. Kubo, M. Kuwabara, J. Yang, Acoustic-cavitation-based production of foamed metallic material, Jpn. J. Appl. Phys. **45** (5B) (2006) 4793–4799, https://doi.org/10.1143/JJAP.45.4793.

[18] K. Oikawa, F. Maekawa, M. Harada, T. Kai, S. Meigo, Y. Kasugai, M. Ooi, K. Sakai, M. Teshigawara, S. Hasegawa, M. Futakawa, Y. Ikeda, N. Watanabe, Design

and application of NOBORU—NeutrOn Beam line for Observation and Research Use at J-PARC, Nucl. Instrum. Meth. Phys. Res. A **589** (2008) 310-317, https://doi.org/10.1016/j.nima.2008.02.019.

[19] M. Belchuk, D. Watt, J. Dryden, Determining a constitutive equation for creep of a Wood's metal model material: constitutive laws of plastic deformation and fracture,19th Canadian Fracture Conference (Ottawa, Ontario, 29–31 May 1989), pp 189-195, https://doi.org/10.1007/978-94-009-1968-6_21.

[20] H. Kamioka, A. Kataoka, Properties of elastic waves in a polycrystalline solid during melting and solidification processes (I)-dilatational wave in Wood's alloy, Zisin **30** (1977) 463-475, https://doi.org/10.4294/zisin1948.30.4_463.

[21] P. Trtik, E. H. Lehmann, Progress in high-resolution neutron imaging at the Paul Scherrer Institut – the neutron microscope project, J. Phys. Conf. Ser. **746** (2016) 012004, https://doi.org/10.1088/1742-6596/746/1/012004.

[22] T. D. Vu, Y. Iizawa, K. Nishimura, H. Shishido, K. M. Kojima, K. Oikawa, M. Harada, S. Miyajima, M. Hidaka, T. Oku, K. Soyama, K. Aizawa, T. Koyama, T. Ishida, Temperature dependent characteristics of neutron signals from a current biased Nb nanowire detector with $^{10}$B converter, J. Phys. Conf. Ser. **1293** (2019) 012051, https://doi.org/10.1088/1742-6596/1293/1/012051.

[23] T. D. Vu, H. Shishido, K. M. Kojima, T. Koyama, K. Oikawa, M. Harada, S. Miyajima, T. Oku, K. Soyama, K. Aizawa, M. Hidaka, S. Y. Suzuki, M. M. Tanaka, A. Malins, M. Machida, T. Ishida, Homogeneity of neutron transmission imaging over a large sensitive area with a four-channel superconducting detector, Supercond. Sci. Technol. 34 (2021) 015010 (10pp), https://doi.org/10.1088/1361-6668/abc2af.

[24] A. Malins, M. Machida, T. D. Vu, K. Aizawa, T. Ishida, Monte Carlo radiation transport modelling of the current-biased kinetic inductance detector, Nucl. Instrum. Meth. Phys. Res. A **953** (2020) 163130, https://doi.org/10.1016/j.nima.2019.163130.

**Fig. 1.** Schematic diagram of the superconducting neutron imaging system. Pulsed neutrons were incident on the substrate side of the detector after passing through the test samples from the 14 m long beam-line at BL10 of the MLF at J-PARC. The detector and samples were cooled down to cryogenic temperatures using a Gifford–McMahon cryocooler. The neutron detector consists of the X and Y-meanderlines, which are superimposed orthogonally to each other, and a $^{10}$B neutron conversion layer. Neutron signals arising from both ends of the two meanderlines were amplified by low noise amplifiers at each channel (Ch1, Ch2, Ch3, or Ch4) to feed into a Kalliope-DC readout circuit and an oscilloscope via four signal splitters. The system was controlled by a data acquisition (DAG) program and LabVIEW software.

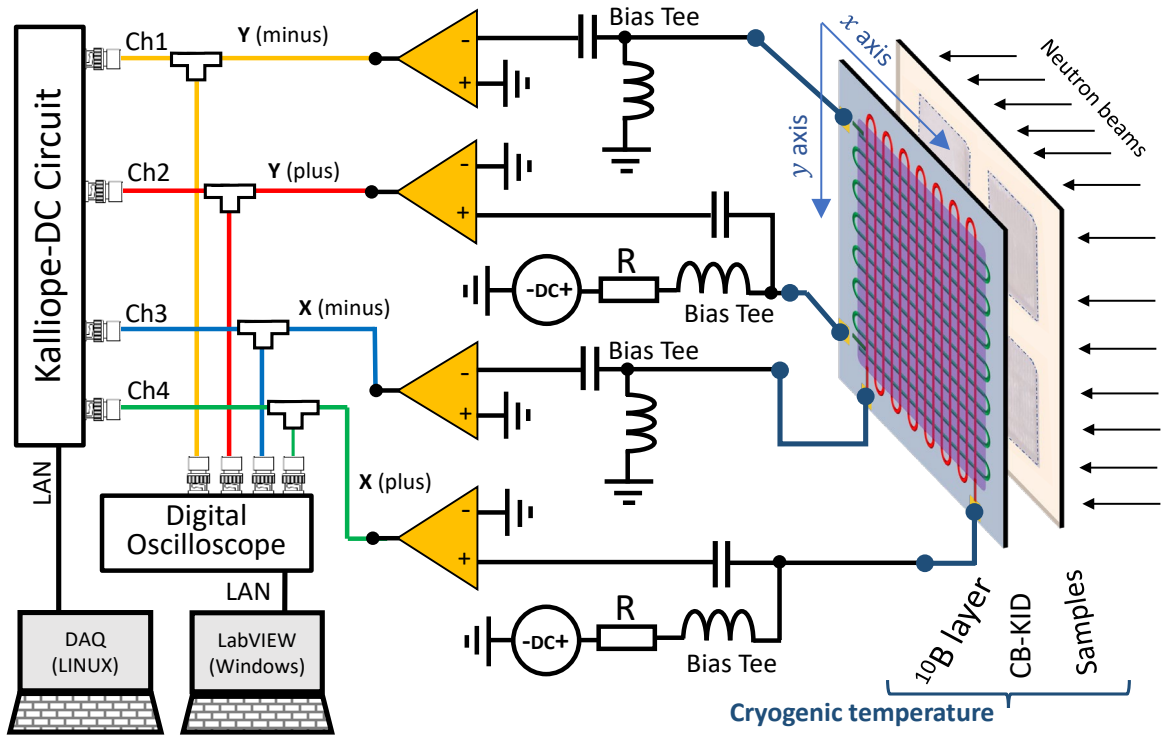

**Fig. 2**. **(a)** Neutron transmission image of the Gd sample on the Si substrate taken using CB-KID with neutron wavelengths 0.1 to 1.13 nm. **(b)** Corresponding SEM image. The Gd sample was dry etched using Ar ion milling.

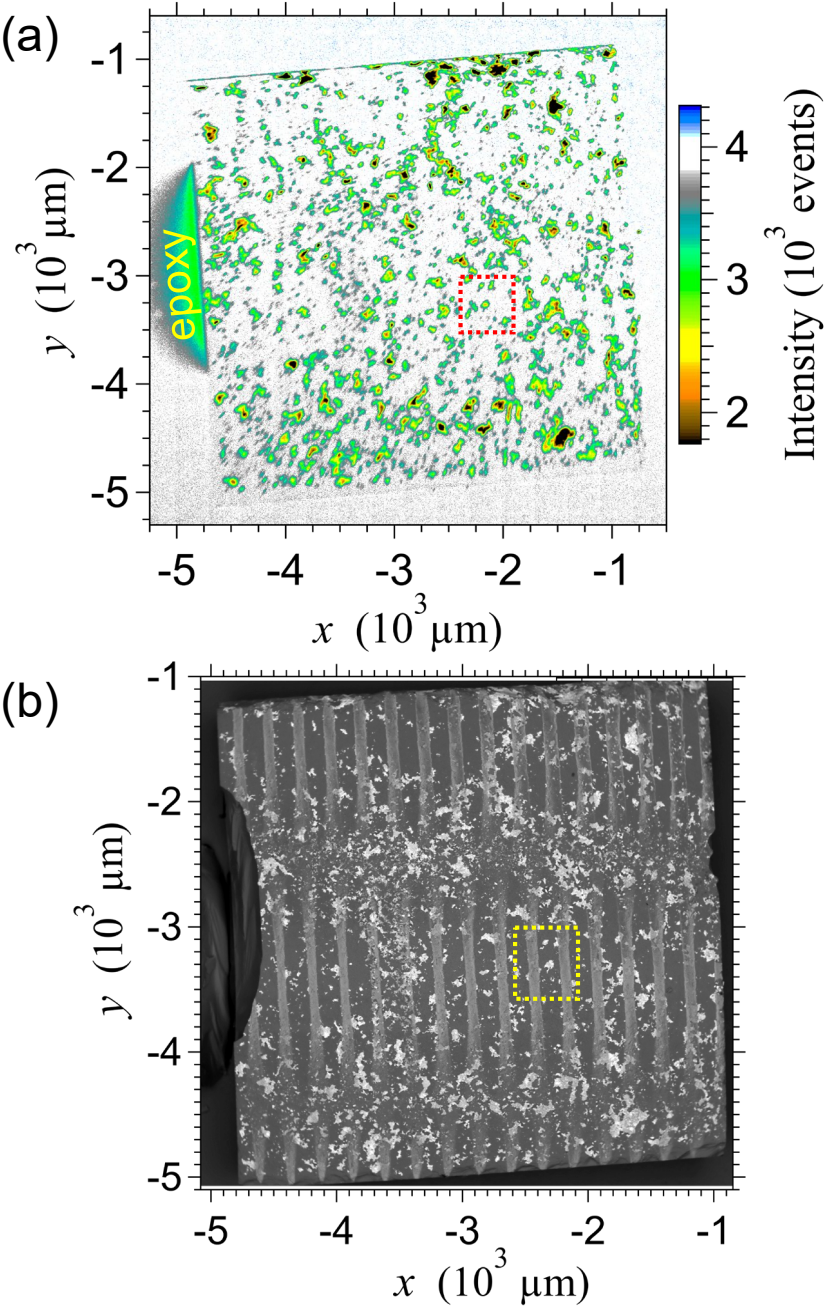

**Fig. 3. (a)** and **(b)** Enlargements of the marked areas in **Fig. 1(a)** (neutron transmission image) and **Fig. 1(b)** (SEM image) of the Gd sample, respectively. **(c)** and **(d)** Line profiles and fitting curves of the Gd island along the *x* direction for the neutron transmission and SEM images, respectively. **(e)** and **(f)** As per above panels, but for line profiles along the *y* direction across the Gd island. **(g)** Correlation of the sizes of the various Gd islands estimated using the neutron transmission and SEM images. The red and green fitting lines overlap with each other in the figure.

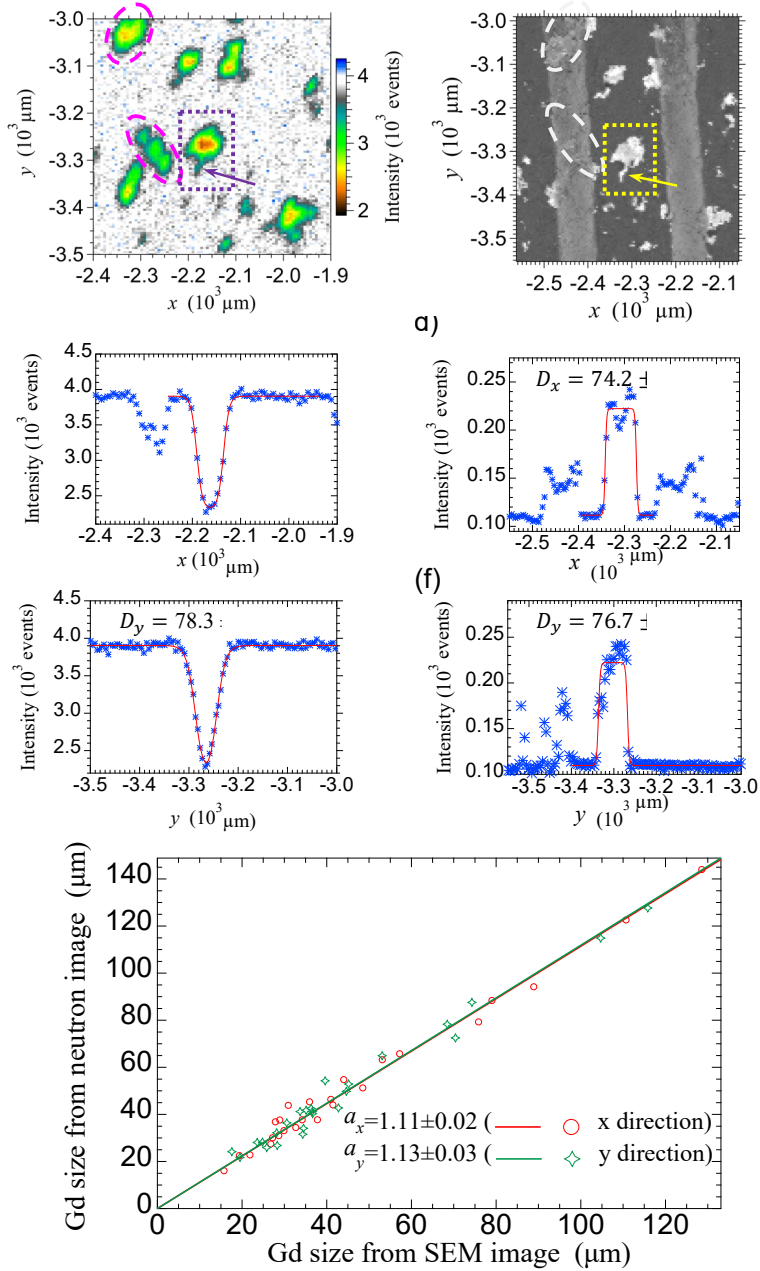

**Fig. 4. (a)** Neutron transmission image of the $Gd_2O_3$ sample on the Si substrate taken with neutron wavelengths 0.1 to 1.13 nm. **(b)** SEM image of the $Gd_2O_3$ sample. This sample was also dry etched using Ar ion milling. **(c)** Enlargement of a void apparent in the neutron transmission image of the $Gd_2O_3$ sample (see dotted area of **(a)**). **(d)** Line profile of the void along the $x$ direction. **(e)** Line profile of a void along the $y$ direction.

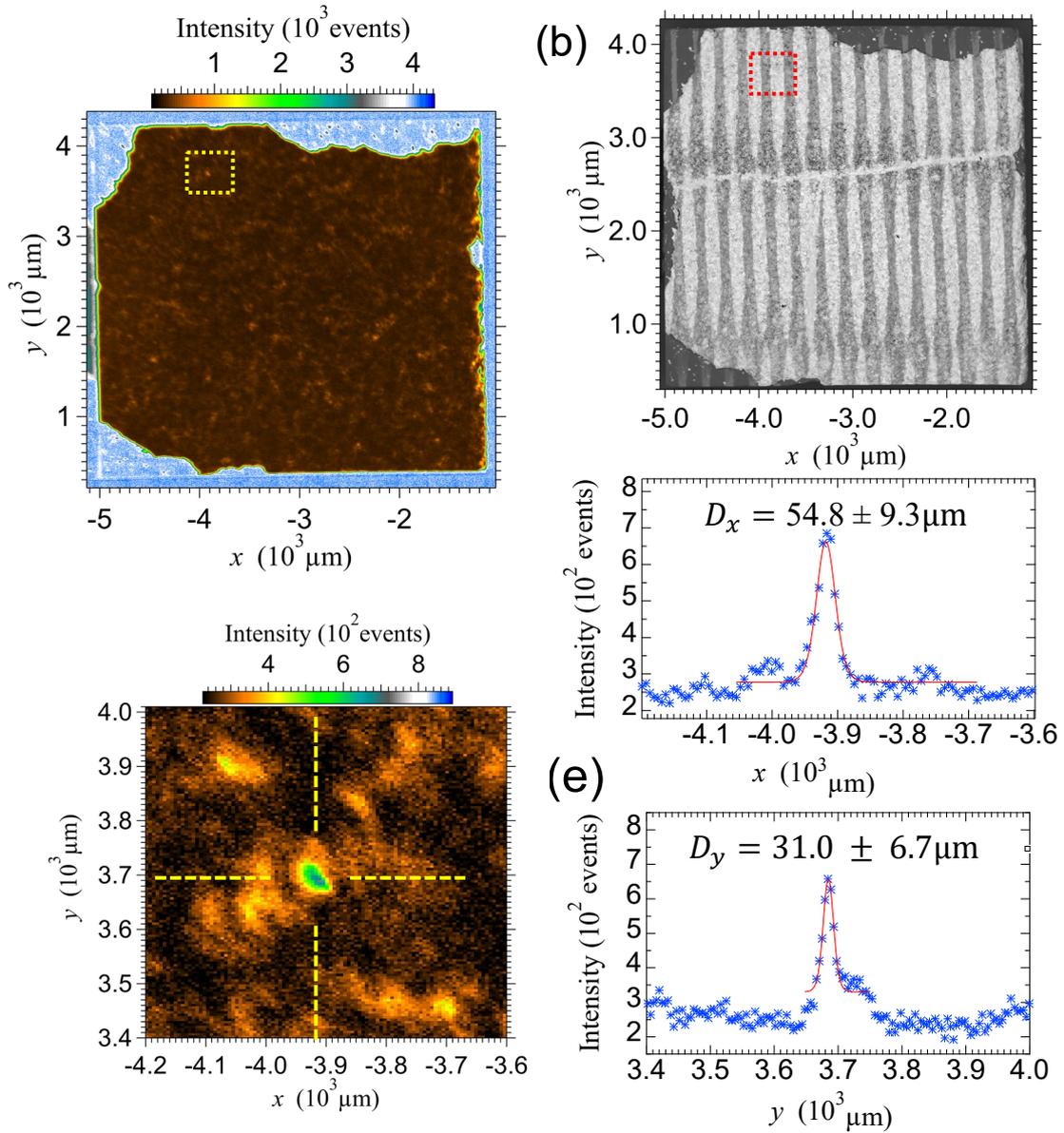

**Fig. 5.** Features in neutron transmission images taken with neutron wavelengths 0.1 to 1.13 nm of Wood's metal samples. The sample thicknesses were (a) 0.24 mm, (b) 0.80 mm, (c) 0.28 mm, (d) 0.39 mm, (e) 0.35 mm and (f) 0.71 mm. Various different patterns can be seen. Dendritic white lines in the images are from Cd-rich phases, while dark lines are from Cd-deficient phases.

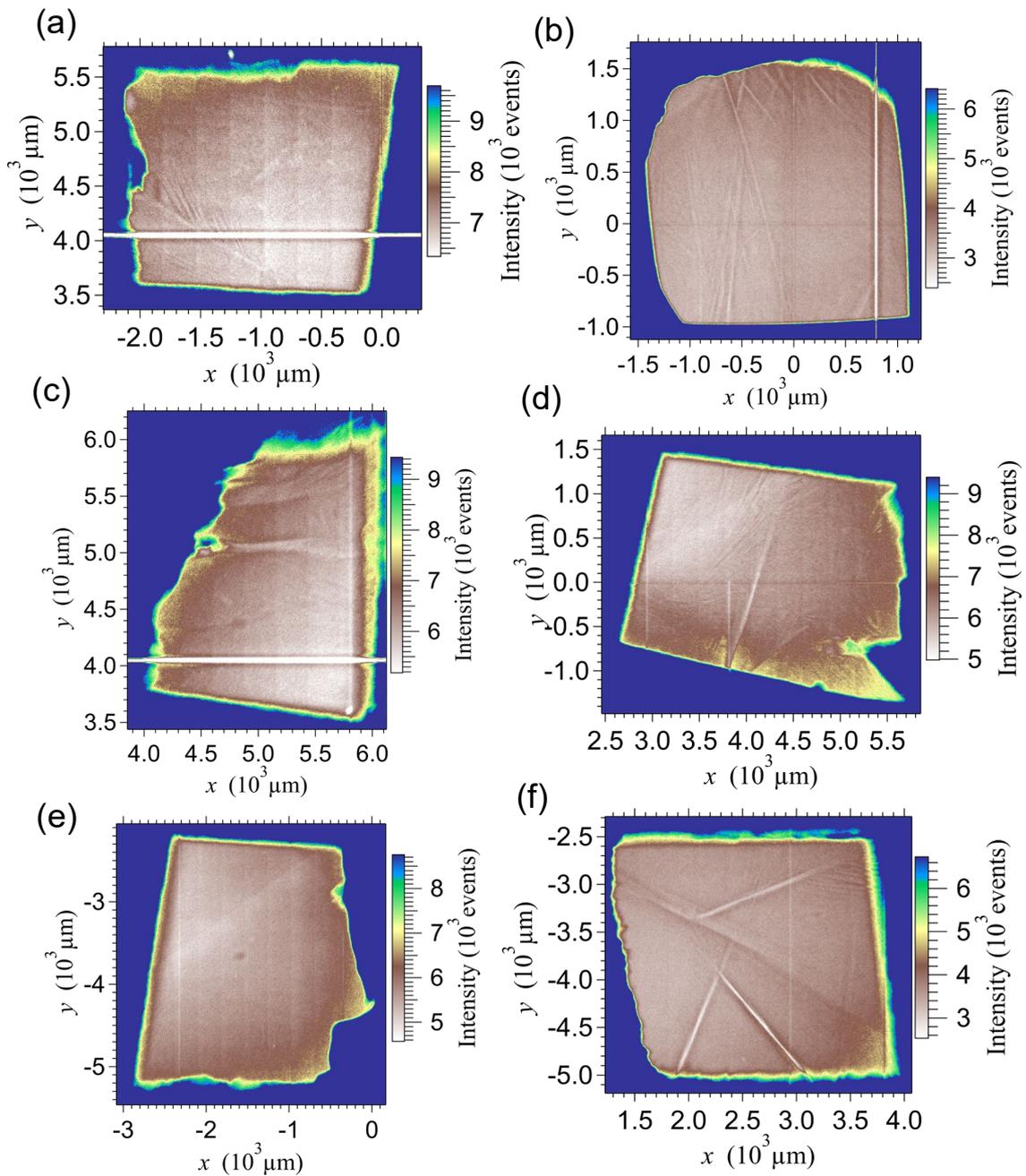

**Fig. 6.** Experimental detection efficiencies on the *X* and *Y* meanderlines separately, and simultaneously on both *X* & *Y* meanderlines with 70-nm thick $^{10}$B conversion layer as a function of neutron wavelength (or time of flight) when the detector temperature was controlled at 7.9 K. Efficiencies calculated from PHITS simulations are also shown (dashed lines).

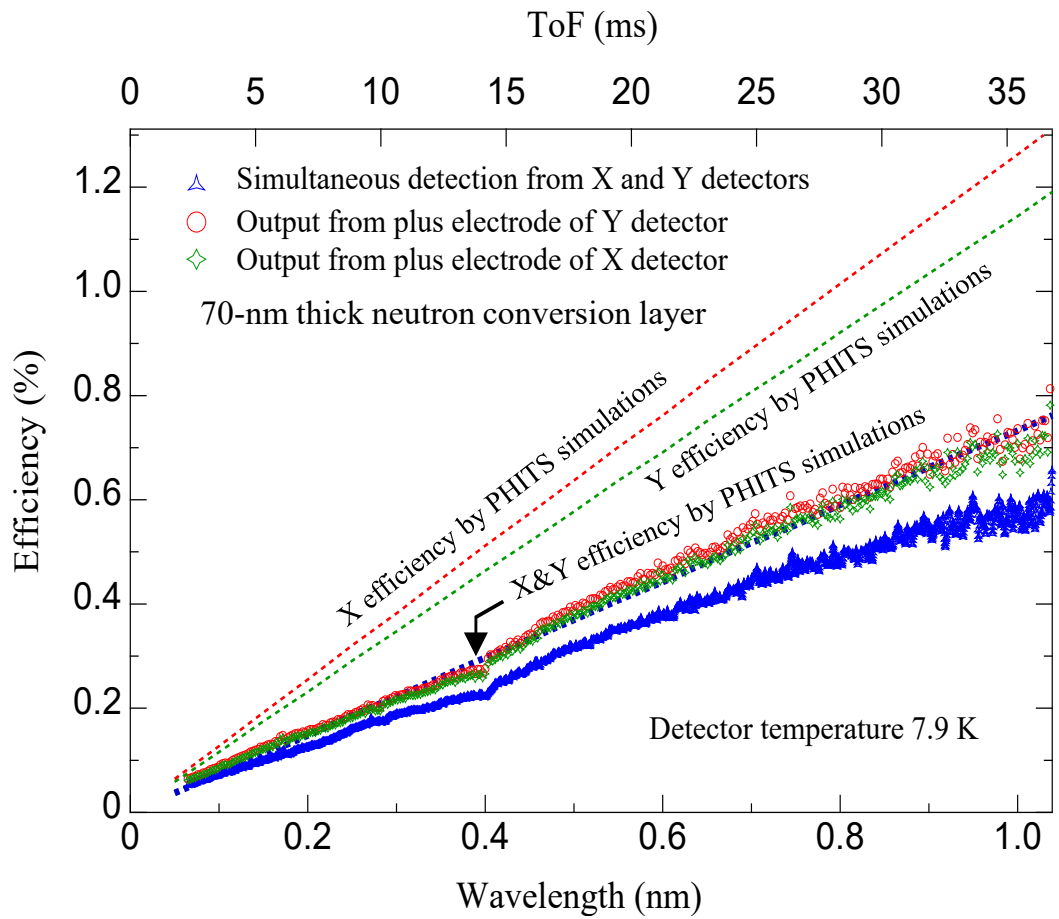